# Energy Efficient virtualization framework for 5G F-RAN

**Yu Zeng, Ahmed Al-Quzweeni, Taisir E.H. Elgorashi, and Jaafar M.H. Elmirghani** *Senior Member, IEEE*

*School of Electronic & Electrical Engineering, University of Leeds, LS2 9JT, United Kingdom*

**ABSTRACT**
Fog radio access network (F-RAN) and virtualisation are promising technologies for 5G networks. In F-RAN, the fog and cloud computing are integrated where the conventional C-RAN functions are diverged to the edge devices of radio access networks. F-RAN is adopted to mitigate the burden of front-haul and improve the end to end (E2E) latency. On other hand, virtualization and network function virtualization (NFV) are IT techniques that aim to convert the functions from hardware to software based functions. Many merits could be brought by the employment of NFV in mobile networks including a high degree of reliability, flexibility and energy efficiency. In this paper, a virtualization framework is introduced for F-RAN to improve the energy efficiency in 5G networks. In this framework, a gigabit passive optical network (GPON) is leveraged as a backbone network for the proposed F-RAN architecture where it connects several evolved nodes B (eNodeBs) via fibre cables. The energy-efficiency of the proposed F-RAN architecture has been investigated and compared with the conventional C-RAN architecture in two different scenarios using mixed integer linear programming (MILP) models. The MILP results indicate that on average a 30% power saving can be achieved by the F-RAN architecture compared with the C-RAN architecture.

**INTRODUCTION**

In 1947, the architecture and concept of cellular communications were proposed for the first time by the American Telephone & Telegraph (AT&T) company [1]. Since then, wireless communication systems have experienced several significant evolutions, achieving transformations from a simplex analogue voice network to a heterogeneous and efficient communication system. The current mobile generation namely "4G" supports many applications and a huge number of users. In the last ten years, with the development of information technology, the advent of new mobile technologies such as high resolution video and Internet of Things (IoT) have transformed wireless communication systems form a network that connects people, to a network of anything at anytime and anywhere [2] – [6]. The rapid growth in the number of connected devices and the rise in the variety and data needs of applications has resulted in an explosive traffic growth in the network and harsh E2E requirements [7]. Therefore, the traffic volume in the next generation of mobile networks (5G) is expected to increase by a factor of 1000 compared to current mobile communication systems [8] – [12] while the latency and other requirements comprehensively transcend the capabilities of 4G communication systems [13] – [15]. C-RAN and NFV were studied as potential 5G solutions which can achieve the target of reducing signal interference at the edge of cellular networks, and support adaptive spectrum slicing and sharing via centralized management and coordination between eNodeBs [16], [17]. In C-RAN and NFV deployments, the function of the base band unit (BBU) can be separated from eNodeBs and can be virtualized to construct BBU pools in the access network (such as GPON access network) to achieve improved network resources sharing. However, due to massive social applications, redundant traffic at front-haul and BBU pools becomes a bottleneck of C-RAN [15], [18].

Information and Communication Technology (ICT) is predicted to account for 2.7% of the global carbon dioxide emissions in 2020 [16], [19] – [21]. Therefore, energy-efficiency is a significant concern in 5G design [22], [23]. From the perspective of commercial profit and environmental targets, the power consumption of 5G mobile networks is targeted to be one tenth of the power consumption of current mobile networks [15], [24] – [25]. The authors of [2], [26] – [29] proposed GPON and IP over WDM as an optical network architecture for virtualization to improve the energy efficiency in 5G networks while the authors of [30] focused in their work on RAN power consumption as it consumes around 70% to 80% of the total mobile energy consumption. F-RAN is proposed as an alternative to 5G RAN. UEs and eNodeBs are endowed with data caching and signal processing capacity in F-RAN architecture, accompanied by virtualized central management [31]. In this work, we introduce an energy-efficient virtualization framework for F-RAN in 5G networks. In this framework the energy-efficiency of the F-RAN architecture is compared with the energy efficiency of the conventional C-RAN architecture using MILP models. In conventional C-RAN, the VMs are hosted by GPON nodes only while in the F-RAN architecture, VMs are hosted by GPON nodes, eNodeBs, and UDs. The rest of this paper is organized as follows: Section 2 introduces and discusses the proposed architecture, Section 3 presents the developed MILP model results, while the conclusions are drawn in Section 4.

**F-RAN architecture**

In F-RAN, data caching and processing capacity of edge devices and nodes (such as UDs and eNodeBs) can be leveraged on a virtualized platform to improve the flexibility and efficiency of RAN [31]. The proposed F-RAN architecture is shown in Fig. 1. An OLT is considered in this architecture connecting two ONUs. Each ONU is connected to one eNodeB. In the service area of each eNodeB, UDs communicate with each other directly using D2D links and indirectly using other UDs, while the virtualized BBU communicate using GPON. VMs in the proposed architecture are considered to carry out the BBU functions such as collaboration radio signal processing (CRSP) and cooperative radio resource management (CRRM) to manage the storage and the processing abilities of edge devices. VMs could be accommodated by GPON nodes, eNodeBs, and UDs, and their location is optimized and migrated based on user demand for the purpose of energy-efficiency. Hence, the requests from users can be directly processed or routed to neighbouring UDs or eNodeBs, in conjunction with the data processing capacity and virtual machined deployments.

We have assumed two different RAN architectures: C-RAN and F-RAN. In C-RAN, VMs are hosted by GPON nodes, while in the F-RAN architecture, VMs are hosted by all elements of F-RAN. A MILP model is developed to investigate the energy efficiency of RANs in 5G networks. The developed MILP model minimizes the total power consumption and optimizes the VM locations associated with the various UDs requests. In order to investigate the energy efficiency of the two RAN architectures, we have invoked two scenarios. In the first scenario, a number of factors such as link capacity, processing capacity of devices and nodes, cycles per instruction, and the power consumption of RAN architectures are tested with various number of active users (load) in a day. In the second scenario, we considered latency requirement of social applications (actual services time). The energy-efficiency of the two architectures are verified under different maximum acceptable latency of requests.

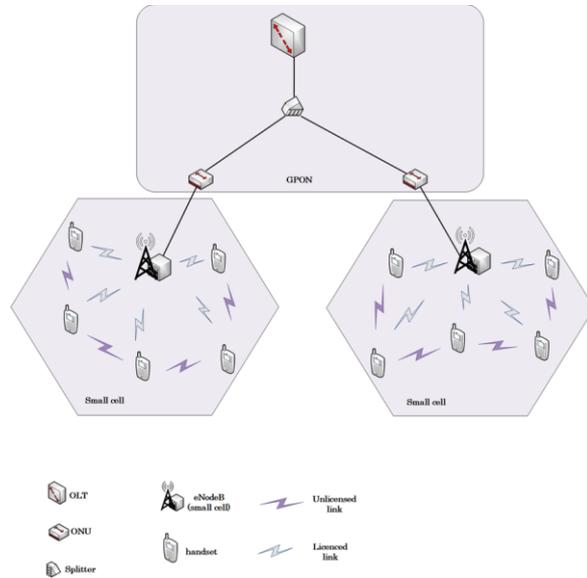

*Figure 1. Proposed F-RAN architecture in 5G*

**RESULTS**

We have considered two groups of users: 11 UDs and 10 UDs separately distributed in the service area of two eNodeBs. In the service area of each eNodeB, UDs are connected with each other via unlicensed channels and connect with eNodeBs via licensed channels. Each UD randomly sends three different requests to the VM with maximum acceptable latency in the mobile network. As alluded to earlier, two scenarios have been considered. In the first scenario the influence of network load is investigated (various number of users) while the second scenario investigates the maximum latency of requests. The network load variation over a day is illustrated in Fig. 2. The power consumption of the two architectures (C-RAN and F-RAN) is compared over different time slots of a day as illustrated in Fig. 3. The power consumption of both architectures is composed of the power consumption of routing traffic and processing the data at GPON, eNodeBs and UDs.

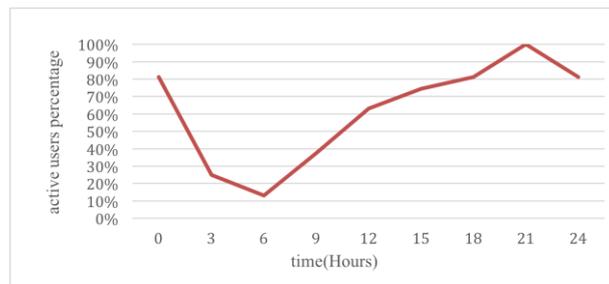

*Figure 2. Active Users Percentage in a Day*

Fig.3 shows that the power consumption of the two architectures fluctuates with the network load over the day; as the total number of active users determines the traffic volume of RAN. The C-RAN architecture has higher power consumption compared with the F-RAN architecture. This is mainly caused by the large number of candidate nodes of F-RAN architecture compared to C-RAN architecture that host VMs user data processing. However, requests from users in F-RAN architecture can be processed by VMs hosted by the closest nodes of the request initiator. The power consumption of traffic traveling around the network experiences a wide variation. Overall, 34% of the power consumption is saved on average with the F-RAN architecture compared to C-RAN architecture.

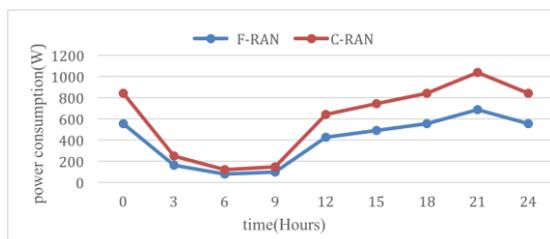 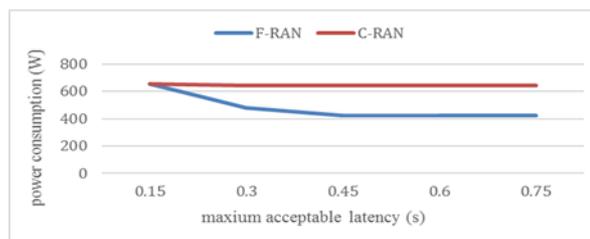

*Figure 3. Power Consumption at Different Times of a Day*  *Figure 4. Power consumption of two architecture with different maximum acceptable latency*

In the second scenario, an M/M/1 queueing model is applied to investigate the maximum latency of the user requests where the processing delay is considered to dominate the propagation delay. According to the M/M/1 queueing model, the delay is expressed as the reciprocal of (the processing capacity of nodes minus the traffic arrival rate). User requests may be processed by multiple nodes. In this case, the latency is determined by the node with maximum latency. We varied the maximum acceptable latency of UDs requests to test the energy efficiency of the two architectures and the results are shown in Fig. 4. In Fig.4, the power consumption of F-RAN is similar to the C-RAN power consumption with a harsh latency restriction. With increase in the maximum acceptable latency, the power consumption of F-RAN steeply diminishes and stays constant with further increase in the acceptable latency, while the power consumption of the C-RAN stays at the same level. In general, the F-RAN architecture has an average power saving of 26% compared to C-RAN. Considering the difference in processing capacity of edge devices and core network nodes, the requests will not be processed at edge devices, when the minimal latency of devices exceeds the maximum acceptable latency of requests. Therefore, driving up the processing capacity of edge devices will further improve the energy efficiency of F-RAN in this scenario.

## CONCLUSIONS

This paper has introduced a virtualization framework for an energy efficient F-RAN architecture in 5G networks. The energy consumption of the proposed F-RAN architecture has been investigated alongside the energy consumption of the conventional C-RAN in two different scenarios. In the first scenario, the energy consumption of the two RAN architectures have been investigated with various network load. The F-RAN architecture shows an average energy saving of 34% compared with C-RAN architecture in conjunction with different network loads. In the second scenario, the influence of maximum acceptable latency of user request has been considered using an M/M/1 queueing model. The energy consumption of two RAN architectures has been studied with different latency values. The F-RAN architecture shows an average power saving of 26% compared to C-RAN architecture. We have found that improving the data processing capacity of edge devices by employing NFV promotes the energy efficiency of F-RAN.

## 5. ACKNOWLEDGEMENTS

The authors would like to acknowledge funding from the Engineering and Physical Sciences Research Council (EPSRC), INTERNET (EP/H040536/1), STAR (EP/K016873/1) and TOWS (EP/S016570/1) projects. All data are provided in full in the results section of this paper.